\documentclass[aps,prab,twocolumn,superscriptaddress,nofootinbib]{revtex4-2}

\usepackage{physics}
\usepackage{amsmath}
\usepackage{amsfonts}
\usepackage{graphicx}
\SetSymbolFont{largesymbols}{bold}{OMX}{txex}{b}{n}
\usepackage{bm}

\newcommand{\mat}[1]{\mathsf{#1}}

\newcommand{\thetamin}{\theta_{\mathrm{min}}}
\newcommand{\thetamax}{\theta_{\mathrm{max}}}
\newcommand{\T}{\mathrm{T}}
\DeclareMathOperator{\diag}{diag}

\begin{document}

\title{Polarization-dependent exchange effects in intrabeam scattering}

\author{Joseph P. Devlin}
\email{Contact author: jpd285@cornell.edu}
\affiliation{Department of Physics, Cornell University, Ithaca, NY 14853, USA}
\author{Georg H. Hoffstaetter}
\affiliation{Department of Physics, Cornell University, Ithaca, NY 14853, USA}
\affiliation{Brookhaven National Laboratory, Upton, NY 11973, USA}

\date{October 26, 2026}

\begin{abstract}
The emittance growth due to intrabeam scattering (IBS) depends on the scattering cross section and is therefore affected by the exchange interaction of identical particles. However, as the IBS growth rates are typically dominated by small-angle scattering events and the exchange interaction only significantly alters the cross section at relatively large angles, the influence of the exchange interaction is usually small. In this paper, we calculate the small corrections to the growth rates resulting from the exchange interaction. We include the polarization dependence of the exchange term in the cross section, and so we also present the growth rates as a function of the beam polarization.
\end{abstract}

\maketitle

\section{Introduction}
The Coulomb interaction within charged-particle beams leads to important collective effects in many modern storage rings. When the beam is modeled as a continuous cloud of charge, a smooth space-charge force arises. However, there are effects which do not appear for a continuous charge distribution, such as those caused by fluctuations of the electromagnetic field on scales comparable to the mean interparticle spacing, i.e., scattering events. 

Scattering within beams leads to two distinct phenomena: the Touschek effect and intrabeam scattering (IBS). The Touschek effect describes the immediate particle loss caused by a single large-angle scattering event which transfers so much momentum from the transverse motion into the longitudinal motion that at least one of the scattered particles ends up outside of the ring's momentum aperture~\cite{Touschek}. These events are possible because, in the rest frame of a high-energy beam, the longitudinal motion is much ``colder" than the transverse motion. IBS, on the other hand, describes the cumulative effect of many small-angle scattering events. The scattered particles are not immediately lost, but the beam size grows. In modern colliders, this growth can be significant during operation and leads to a reduction in luminosity.

As IBS involves relatively small momentum transfers, it is effectively described by the Fokker-Planck equation~\cite{Nash-IBS}. We will not attempt to solve this equation directly, but we will calculate its coefficients, namely the damping vector $b_i$ and the symmetric diffusion tensor $d_{ij}$. We use the definitions
\begin{subequations}\label{eq:damping-and-diffusion}
\begin{align}
\label{eq:damping}
b_i&=\overline{\frac{\delta p_i}{\delta t}},
\\
\label{eq:diffusion}
d_{ij}&=\overline{\frac{\delta p_i\delta p_j}{\delta t}},
\end{align}
\end{subequations}
where $\delta\vec{p}$ is the total momentum transfer resulting from scattering in a time $\delta t$ and the bar indicates averaging over the stochastic sequence of collisions~\cite[\S21]{LP-Kinetics}. These coefficients are generally functions of phase space, and for the case of IBS, they also depend on the beam's phase-space distribution. 

The most well-known theoretical treatments of IBS were published by Piwinski in Ref.~\cite{Piwinski} and by Bjorken and Mtingwa in Ref.~\cite{Bjorken-Mtingwa}. Their calculations do not involve the Fokker-Planck coefficients because they directly compute the growth rate of the beam's second moments, defined by
\begin{equation}
g_{ij}=\dv{t}\langle p_ip_j\rangle,
\end{equation}
where angular brackets indicate averaging over the phase-space distribution. Only IBS-induced growth is included in $g_{ij}$, i.e., growth from single-particle dynamics or from other collective effects is neglected. Additionally, $g_{ij}$ is the growth rate at a single point in the lattice; one usually averages this quantity over the lattice. Sometimes, $g_{ij}$ is called the momentum diffusion rate (e.g., in Ref.~\cite{Kubo-Oide}), but it should not be confused with the single-particle diffusion tensor defined by Eq.~\eqref{eq:diffusion}. The connection between the Fokker-Planck coefficients and the growth rates is~\cite{Nash-Thesis}
\begin{equation}
\label{eq:growth-rates-FP}
g_{ij}=\langle b_ip_j\rangle + \langle b_jp_i\rangle + \langle d_{ij}\rangle.
\end{equation}

The effects of IBS were significant in the Relativistic Heavy Ion Collider (RHIC)~\cite{RHIC-IBS} and are expected to be important for the Hadron Storage Ring (HSR) of the Electron-Ion Collider (EIC)~\cite{HSR-IBS}. These accelerators are special because they stored (or will store) proton beams with a design polarization around ${70\%}$. The polarization $P$ of a beam containing $N$ particles is defined as
\begin{equation}
P=\norm{\frac{1}{N}\sum_{i=1}^N\vec{S}_i},
\end{equation}
where $\vec{S}_i$ is the unit-normalized, rest-frame spin expectation value for particle $i$~\cite{Georg-Book}. It is widely known that the Touschek lifetime depends on the beam polarization~\cite{Lee-Touschek,Fu-Touschek}, but the polarization dependence of IBS growth rates is not often acknowledged~\cite{Zimmermann-IBS}. This effect is generally expected to be small as IBS is dominated by small-angle scattering and the effect of polarization is to modify the exchange term in the cross section, which is only significant at large angles and is therefore usually neglected~\cite{Kabel-Exchange}. Nevertheless, it is beneficial to quantify the polarization dependence and we undertake such a calculation in this paper.

\section{Calculations}
\subsection{Collision rates}
Consider a beam with $N$ particles of mass $m$ distributed according to a unit-normalized phase-space density $\rho$. If the motion of particles within the beam is assumed to be non-relativistic, the collision rate for a particle with position $\vec{x}$ and momentum $\vec{p}$ is given by
\begin{equation}
\label{eq:NR-dn}
\dd{n}=\frac{N}{m}\|\vec{p}-\vec{p}_s\|\rho(\vec{x},\vec{p}_s)\dd[3]{\vec{p}_s}\dd{\sigma}\dd{t},
\end{equation}
where $\vec{p}_s$ is the momentum of the other colliding particle and $\dd{\sigma}$ is the differential cross section of the interaction~\cite[\S41]{LP-Kinetics}. We can use Eq.~\eqref{eq:NR-dn} to find the damping vector and the diffusion tensor:
\begin{subequations}
\begin{align}
b_i(\vec{x},\vec{p})&=\frac{N}{m}\int\dd[3]{\vec{p}_s}\int\dd{\sigma}\|\vec{p}-\vec{p}_s\|\rho(\vec{x},\vec{p}_s)\delta p_i,
\\
d_{ij}(\vec{x},\vec{p})&=\frac{N}{m}\int\dd[3]{\vec{p}_s}\int\dd{\sigma}\|\vec{p}-\vec{p}_s\|\rho(\vec{x},\vec{p}_s)\delta p_i\delta p_j,
\end{align}
\end{subequations}
where $\delta \vec{p}$ is the change in momentum due to a collision. As the relative momentum ${\vec{u}=\vec{p}-\vec{p}_s}$ appears many times, we will use it as our integration variable:
\begin{subequations}\label{eq:NR-damping-and-diffusion}
\begin{align}
\label{eq:NR-damping}
b_i(\vec{x},\vec{p})&=\frac{N}{m}\int\dd[3]{\vec{u}}\int\dd{\sigma}u\rho(\vec{x},\vec{p}-\vec{u})\delta p_i,
\\
d_{ij}(\vec{x},\vec{p})&=\frac{N}{m}\int\dd[3]{\vec{u}}\int\dd{\sigma}u\rho(\vec{x},\vec{p}-\vec{u})\delta p_i\delta p_j.
\label{eq:NR-diffusion}
\end{align}
\end{subequations}

\subsection{The beam frame}
The motion of particles in accelerators usually cannot be treated as non-relativistic because the particles have very large longitudinal velocities. However, if we boost to a frame moving with the mean longitudinal velocity, the motion is often non-relativistic to a good approximation. We call this frame the \emph{beam frame} in contrast to the laboratory (lab) frame. In both the lab frame and the beam frame, we use the basis ${(\vec{e}_x,\vec{e}_y,\vec{e}_s)}$ where $\vec{e}_x$ is horizontal, $\vec{e}_y$ is vertical, and $\vec{e}_s$ is longitudinal.
The transformation from the beam frame to the lab frame is a Lorentz boost with velocity ${-v\vec{e}_s}$ and corresponding Lorentz factor $\gamma$. 

We would like to compute the damping vector and the diffusion tensor in the beam frame to avoid having to consider relativistic effects. Therefore, we need to know how these objects transform between the beam frame and the lab frame. First, the number of collisions is invariant, so the collision rate in the lab frame is a factor of $\gamma$ smaller due to time dilation. Second, in accordance with accelerator physics customs, we would like to describe the momenta in the lab frame by ${p_{x,y}/p_0}$ and ${\delta=(p-p_0)/p_0}$, where the normalizing factor $p_0$ is the ``reference momentum". We now make two assumptions. The first is that the reference momentum is approximately equal to the average longitudinal momentum of the beam so that ${p_0 \approx m\gamma v}$. The second is that the transverse momenta are small in the lab frame so that ${p_\mathrm{lab} \approx p_{\mathrm{lab},s}}$ (the paraxial approximation). After using the Lorentz transformation for $p_s$ and assuming $\gamma \approx 1$ in the beam frame, we get ${p_{\mathrm{beam},s} \approx (p_0/\gamma)\delta_\mathrm{lab}}$. Finally, we note that the longitudinal boost leaves the transverse momenta invariant. Combining these observations gives the transformation rules
\begin{widetext}
\begin{subequations}\label{eq:damping-and-diffusion-transformation}
\begin{align}
\begin{pmatrix}
b_x
\\
b_y
\\
b_\delta
\end{pmatrix}_\mathrm{lab}
&=
\begin{pmatrix}
\gamma^{-1}p_0^{-1}b_x
\\
\gamma^{-1}p_0^{-1}b_y
\\
\phantom{\gamma^{-1}}p_0^{-1}b_s
\end{pmatrix}_\mathrm{beam},
\\
\begin{pmatrix}
d_{xx} & d_{xy} & d_{x\delta} 
\\
d_{yx} & d_{yy} & d_{y\delta}
\\
d_{\delta x} & d_{\delta y} & d_{\delta\delta}
\end{pmatrix}_\mathrm{lab}
&=
\begin{pmatrix}
\gamma^{-1}p_0^{-2}d_{xx} & \gamma^{-1}p_0^{-2}d_{xy} & \phantom{\gamma}p_0^{-2}d_{xs} 
\\
\gamma^{-1}p_0^{-2}d_{yx} & \gamma^{-1}p_0^{-2}d_{yy} & \phantom{\gamma}p_0^{-2}d_{ys}
\\
\phantom{\gamma^{-1}}p_0^{-2}d_{sx} & \phantom{\gamma^{-1}}p_0^{-2}d_{sy} & \gamma p_0^{-2}d_{ss}
\end{pmatrix}_\mathrm{beam}.
\end{align}
\end{subequations}
The growth rates can also be transformed directly:
\begin{equation}
\label{eq:gr-transformation}
\begin{pmatrix}
g_{xx} & g_{xy} & g_{x\delta} 
\\
g_{yx} & g_{yy} & g_{y\delta}
\\
g_{\delta x} & g_{\delta y} & g_{\delta\delta}
\end{pmatrix}_\mathrm{lab}
=
\begin{pmatrix}
\gamma^{-1}p_0^{-2}g_{xx} & \gamma^{-1}p_0^{-2}g_{xy} & \phantom{\gamma}p_0^{-2}g_{xs} 
\\
\gamma^{-1}p_0^{-2}g_{yx} & \gamma^{-1}p_0^{-2}g_{yy} & \phantom{\gamma}p_0^{-2}g_{ys}
\\
\phantom{\gamma^{-1}}p_0^{-2}g_{sx} & \phantom{\gamma^{-1}}p_0^{-2}g_{sy} & \gamma p_0^{-2}g_{ss}
\end{pmatrix}_\mathrm{beam}.
\end{equation}
\end{widetext}

To proceed, we calculate the damping vector and the diffusion tensor in the beam frame. Then, we combine these quantities into growth rates. Finally, we transform the growth rates to the lab frame.

\subsection{Damping vector and diffusion tensor in the beam frame}
We begin with Eq.~\eqref{eq:NR-damping-and-diffusion} and determine $\delta\vec{p}$ in terms of the appropriate quantities. We will express the cross section $\dd{\sigma}$ in terms of the solid-angle element ${\dd{\Omega}=\sin\theta\dd{\theta}\dd{\phi}}$, and so $\delta \vec{p}$ must be determined as a function of $\vec{u}$ and the spherical angles $(\theta,\phi)$. For this calculation, it is convenient to use an orthonormal basis $(\vec{e}_1,\vec{e}_2,\vec{e}_3)$ where ${\vec{u}=u\vec{e}_3}$. Then, the relative momentum is rotated by the collision:
\begin{equation}
\delta\vec{u}=u\qty[(\vec{e}_1\cos\phi+\vec{e}_2\sin\phi)\sin\theta+\vec{e}_3(\cos\theta-1)],
\end{equation}
and conservation of momentum implies ${\delta\vec{u}=2\delta\vec{p}}$. 

For many interactions, including those considered here, the cross section is independent of the azimuthal angle $\phi$. In this case, the relevant quantity is the integral of $\delta\vec{p}$ over $\phi$:
\begin{widetext}
\begin{subequations}\label{eq:momentum-changes}
\begin{align}
\label{eq:momentum-change-a}
\int_0^{2\pi}\dd{\phi}\delta p_i&=\pi u_i(\cos\theta-1),
\\
\label{eq:momentum-change-b}
\int_0^{2\pi}\dd{\phi}\delta p_i\delta p_j&=\frac{\pi}{4}\qty(u^2\delta_{ij}-u_iu_j)\sin^2\theta+\frac{\pi}{2}u_iu_j(\cos\theta-1)^2.
\end{align}
\end{subequations}
Hence, the damping vector and the diffusion tensor are
\begin{subequations}\label{eq:damping-and-diffusion-before-cross-section}
\begin{align}
\label{eq:damping-before-cross-section}
b_i(\vec{x},\vec{p})&=\frac{\pi N}{m}\int\dd[3]{\vec{u}}u_iu \rho(\vec{x},\vec{p}-\vec{u})\int_{\thetamin}^{\thetamax}\dd{\theta}\sin\theta(\cos\theta-1)\dv{\sigma}{\Omega},
\\
\label{eq:diffusion-before-cross-section}
d_{ij}(\vec{x},\vec{p})&=\frac{\pi N}{2m}\int\dd[3]{\vec{u}}u \rho(\vec{x},\vec{p}-\vec{u})\int_{\thetamin}^{\thetamax}\dd{\theta}\sin\theta\qty[\frac{1}{2}\qty(u^2\delta_{ij}-u_iu_j)\sin^2\theta+u_iu_j(\cos\theta-1)^2]\dv{\sigma}{\Omega}.
\end{align}
\end{subequations}
\end{widetext}
Now, we must choose the appropriate cross section. The Piwinski and Bjorken-Mtingwa calculations use the Rutherford cross section for non-relativistic scattering of point particles. In our notation, it takes the form
\begin{equation}
\label{eq:Rutherford}
\dv{\sigma}{\Omega}=\frac{\kappa^2}{u^4\sin^4\frac{\theta}{2}},
\end{equation}
where ${\kappa=(mc)^2r_0}$ for particles with mass $m$ and classical radius $r_0$~\cite[\S19]{LL-Mechanics}. In the resulting calculation, the maximum scattering angle $\thetamax$ can be set to $\pi$, but the integrals diverge logarithmically for ${\thetamin \to 0}$, which is the source of the \emph{Coulomb logarithm}. 

Inserting Eq.~\eqref{eq:Rutherford} into Eq.~\eqref{eq:damping-and-diffusion-before-cross-section}, retaining only the logarithmically divergent terms, and pulling them outside of the integral (i.e., assuming they are independent of $\vec{u}$) leads to the Bjorken-Mtingwa results. We refer to this procedure, in which only logarithmically divergent terms are retained, as restricting the calculation to ``leading logarithmic accuracy". To go past leading logarithmic accuracy, one must take into account that the Rutherford cross section is not appropriate at large angles because it does not respect the indistinguishability of identical particles.

The cross section for non-relativistic scattering of \emph{identical} spin-1/2 point particles in the Born approximation is~\cite[\S137]{LL-Quantum}
\begin{equation}
\label{eq:nonrelativistic-Moller}
\dv{\sigma}{\Omega}=\frac{\kappa^2}{u^4}\qty(\frac{1}{\sin^4\frac{\theta}{2}}+\frac{1}{\cos^4\frac{\theta}{2}}-\frac{1}{\sin^2\frac{\theta}{2}\cos^2\frac{\theta}{2}}).
\end{equation}
Each term in the brackets has a clear meaning: the first term corresponds to the Rutherford cross section, the second term is a classical correction due to the indistinguishability of scattering into supplementary angles, and the third term represents the quantum exchange interaction. When using a cross section for identical particles in Eq.~\eqref{eq:damping-and-diffusion-before-cross-section}, the maximum scattering angle $\thetamax$ must be set to $\pi/2$, or else the final states will be overcounted. The effect of using the Born approximation, and thereby excluding Coulomb effects, is analyzed for the Touschek lifetime in Ref.~\cite{Strakhovenko}.

Even in the non-relativistic limit, the cross section depends on the polarization of the beam. This effect is not due to a direct spin-orbit interaction, but rather that fermions in a symmetric spin state must be in an antisymmetric orbital state, and vice versa. For a beam with polarization $P$, the result is~\cite[\S81]{BLP-QED}
\begin{equation}
\label{eq:polarized-nonrelativistic-Moller}
\dv{\sigma}{\Omega}=\frac{4\kappa^2}{u^4}\qty(\frac{4}{\sin^4\theta}-\frac{3+P^2}{\sin^2\theta}).
\end{equation}
This formula has already been utilized in calculating the polarization dependence of the Touschek lifetime in electron storage rings~\cite{Lee-Touschek,Fu-Touschek}. As one would expect, this cross section is in agreement with Eq.~\eqref{eq:nonrelativistic-Moller} for an unpolarized beam ($P=0$).  

We are now prepared to integrate over $\theta$ in Eq.~\eqref{eq:damping-and-diffusion-before-cross-section}. Before doing so, we note that it is still necessary to impose a minimum scattering angle $\thetamin>0$. This divergence appears because our calculations do not take into account the variation of the beam profile. For example, any real beam has a finite size, so there is a maximum impact parameter and hence a minimum scattering angle. One method of avoiding this divergence is developed in Ref.~\cite{Nash-Thesis}, but we will not treat this topic any further.

\begin{widetext}
We now integrate over $\theta$~\cite{Mathematica}:
\begin{subequations}\label{eq:theta-integrals}
\begin{align}
\label{eq:theta-integral-a}
-\int_{\thetamin}^{\pi/2}\dd{\theta}\sin\theta(\cos\theta-1)\qty(\frac{4}{\sin^4\theta}-\frac{3+P^2}{\sin^2\theta})&=2\ln(\cot\frac{\thetamin}{2})-(3+P^2)\ln(1+\cos\thetamin)-\sec^2\frac{\thetamin}{2}+2,
\\
\label{eq:theta-integral-b}
\frac{1}{2}\int_{\thetamin}^{\pi/2}\dd{\theta}\sin^3\theta\qty(\frac{4}{\sin^4\theta}-\frac{3+P^2}{\sin^2\theta})&=2\ln(\cot\frac{\thetamin}{2})-\frac{3+P^2}{2}\cos\thetamin,
\\
\label{eq:theta-integral-c}
\int_{\thetamin}^{\pi/2}\dd{\theta}\sin\theta(\cos\theta-1)^2\qty(\frac{4}{\sin^4\theta}-\frac{3+P^2}{\sin^2\theta})&=(3+P^2)\qty[\ln(\frac{1}{4}\sec^4\frac{\thetamin}{2})+\cos\thetamin]+\frac{4\cos\thetamin}{1+\cos\thetamin}.
\end{align}
\end{subequations}
\end{widetext}
The first two integrals contain the same logarithmically divergent term. Its asymptotic form for ${\thetamin \to 0}$ is
\begin{equation}
\ln(\cot\frac{\thetamin}{2}) \sim \ln(\frac{2}{\thetamin}),
\end{equation}
which agrees with the Coulomb logarithm ${\Lambda=\ln(2/\thetamin)}$ defined by Bjorken and Mtingwa. The third integral, which contributes to the diffusion tensor along with the second integral, has a finite limit for ${\thetamin \to 0}$ and therefore does not appear when calculating to leading logarithmic accuracy. In our treatment, we retain the non-divergent terms but use their ${\thetamin \to 0}$ limit as the cutoff is not strictly necessary for these terms and they converge to this limit very rapidly for typical values of $\thetamin \ll 1$.

\begin{figure*}
\centering
\includegraphics[scale=0.65]{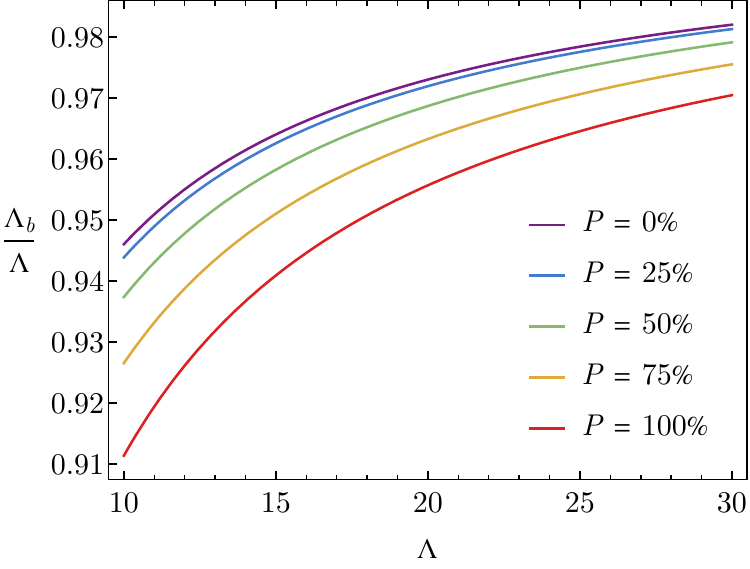}
\includegraphics[scale=0.65]{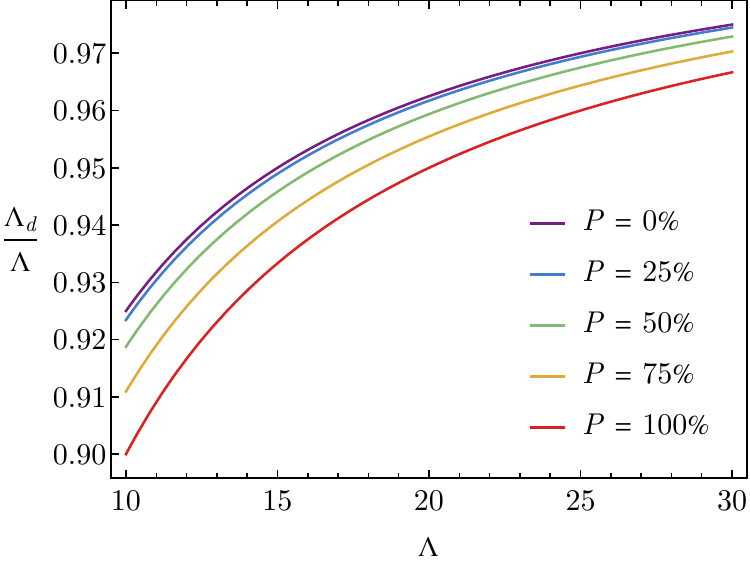}
\includegraphics[scale=0.65]{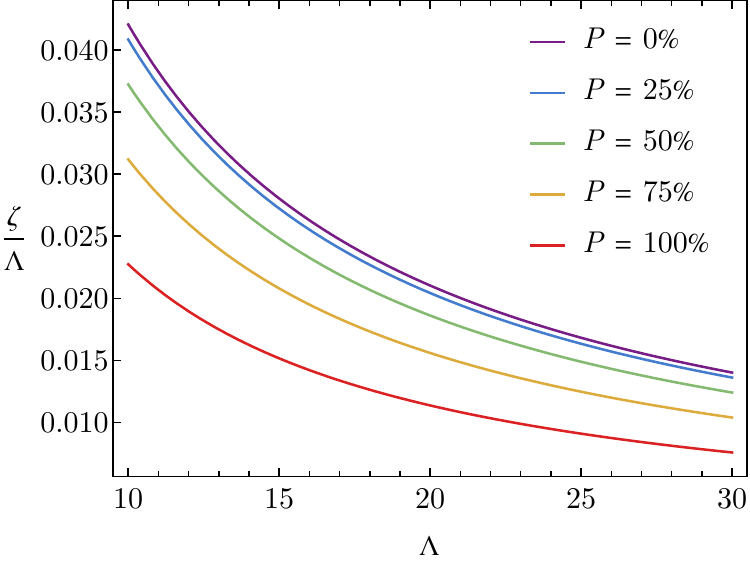}
\caption{\label{fig:LogPlots}Comparison of the logarithmically divergent coefficients $\Lambda_b$ and $\Lambda_d$ and the non-divergent coefficient $\zeta$ to the Coulomb logarithm $\Lambda$ for different polarization levels.}
\end{figure*}

\begin{widetext}
We now insert Eq.~\eqref{eq:theta-integrals} into Eq.~\eqref{eq:damping-and-diffusion-before-cross-section}:
\begin{subequations}\label{eq:damping-and-diffusion-before-distribution}
\begin{align}
\label{eq:damping-before-distribution}
b_i(\vec{x},\vec{p})&=-\frac{8\pi N\kappa^2\Lambda_b}{m}\int\dd[3]{\vec{u}}\frac{u_i}{u^3}\rho(\vec{x},\vec{p}-\vec{u}),
\\
\label{eq:diffusion-before-distribution}
d_{ij}(\vec{x},\vec{p})&=\phantom{-}\frac{4\pi N\kappa^2\Lambda_d}{m}\qty[\int\dd[3]{\vec{u}}\frac{u^2\delta_{ij}-u_iu_j}{u^3}\rho(\vec{x},\vec{p}-\vec{u})+\frac{\zeta}{\Lambda_d}\int\dd[3]{\vec{u}}\frac{u_iu_j}{u^3}\rho(\vec{x},\vec{p}-\vec{u})],
\end{align}
\end{subequations}
\end{widetext}
where we have defined the coefficients
\begin{subequations}\label{eq:logs-and-nonlog}
\begin{align}
\label{eq:logb}
\Lambda_b&=\ln(\cot\frac{\thetamin}{2})-\frac{3+P^2}{2}\ln 2+\frac{1}{2},
\\
\label{eq:logd}
\Lambda_d&=\ln(\cot\frac{\thetamin}{2})-\frac{3+P^2}{4},
\\
\label{eq:nonlog}
\zeta&=\frac{3+P^2}{2}\qty(1-\ln 4)+1.
\end{align}
\end{subequations}
We assume that $\thetamin$ is a constant, i.e., it is independent of $\vec{x}$, $\vec{p}$, and $\vec{u}$. The quantities $\Lambda_b$ and $\Lambda_d$ are analogous to the Coulomb logarithm $\Lambda$, and to leading logarithmic accuracy, ${\Lambda_b=\Lambda_d=\Lambda}$. As mentioned before, $\zeta$ is not logarithmically divergent and therefore does not appear in the work of Bjorken and Mtingwa. A comparison of these three coefficients to the Coulomb logarithm of Bjorken and Mtingwa is shown in Fig.~\ref{fig:LogPlots}. In short, the effect of using the cross section for a beam of polarized fermions [Eq.~\eqref{eq:polarized-nonrelativistic-Moller}] in place of the Rutherford cross section [Eq.~\eqref{eq:Rutherford}] is to modify the logarithmic coefficients of the damping vector and the diffusion tensor, thereby making them unequal ${(\Lambda_b \neq \Lambda_d)}$, and also to add a non-logarithmic term to the diffusion tensor.

To proceed further, it is necessary to specify the distribution $\rho$. The usual choice is a Gaussian:
\begin{equation}
\label{eq:bf-Gaussian}
\rho(\vec{x},\vec{p})=\frac{1}{\Gamma}\exp(-\frac{1}{2}\vec{x}^\T\mat{A}\vec{x}-\vec{x}^\T\mat{B}\vec{p}-\frac{1}{2}\vec{p}^{\,\T}\mat{C}\vec{p}),
\end{equation}
where $\Gamma$ is the total phase-space volume (we only consider bunched beams). For simplicity, we assume that $\vec{x}$ and $\vec{p}$ are measured relative to their mean values. As long as no direction has vanishing momentum spread, it is possible to find a matrix ${\mat{R}\in\mathrm{SO}(3)}$ such that ${\mat{R}\mat{C}\mat{R}^\T=\diag(c_1^{-1},c_2^{-1},c_3^{-1})}$. The momentum in the rotated frame ${(\vec{w}=\mat{R}\vec{p})}$ then has a simpler distribution function. A similar idea was used in Ref.~\cite{Kubo-Oide} to generalize the calculations of Bjorken and Mtingwa to arbitrarily coupled distributions. However, we will make one final transformation to the momentum before continuing:
\begin{equation}
\vec{\xi}=\mat{R}\qty(\vec{p}+\mat{C}^{-1}\mat{B}^\T\vec{x}).
\end{equation}
The resulting distribution is
\begin{equation}
\label{eq:diagonalized-Gaussian}
\tilde{\rho}(\vec{x},\vec{\xi})=\frac{1}{\Gamma}\exp(-\frac{1}{2}\vec{x}^\T\tilde{\mat{A}}\vec{x}-\sum_{k=1}^3\frac{\xi_k^2}{2c_k}),
\end{equation}
where $\tilde{\mat{A}}=\mat{A}-\mat{B}\mat{C}^{-1}\mat{B}^\T$. We have not only diagonalized the momentum-momentum terms, but we have also eliminated the momentum-position terms. We write this distribution in the product form ${\tilde{\rho}(\vec{x},\vec{\xi})=\tilde{\rho}_x(\vec{x})\tilde{\rho}_\xi(\vec{\xi})}$, where
\begin{subequations}\label{eq:product-distributions}
\begin{align}
\label{eq:x-distribution}
\tilde{\rho}_x(\vec{x})&=\frac{1}{\tilde{\Gamma}_x}\exp(-\frac{1}{2}\vec{x}^\T\tilde{\mat{A}}\vec{x}),
\\
\label{eq:xi-distribution}
\tilde{\rho}_\xi(\vec{\xi})&=\frac{1}{\tilde{\Gamma}_\xi}\exp(-\sum_{k=1}^3\frac{\xi_k^2}{2c_k}),
\end{align}
\end{subequations}
with $\tilde{\Gamma}_\xi=\sqrt{(2\pi)^3c_1c_2c_3}$ and $\tilde{\Gamma}_x=\Gamma/\tilde{\Gamma}_\xi$.

\begin{widetext}
In terms of $\vec{\xi}$, the damping vector and the diffusion tensor of Eq.~\eqref{eq:damping-and-diffusion-before-distribution} become
\begin{subequations}\label{eq:damping-and-diffusion-after-completing-square}
\begin{align}
\label{eq:damping-after-completing-square}
\tilde{b}_i(\vec{x},\vec{\xi})&=-\frac{8\pi N\kappa^2\Lambda_b}{m\tilde{\Gamma}_\xi}\tilde{\rho}_x(\vec{x})\int\dd[3]{\vec{u}}\frac{u_i}{u^3}\prod_{k=1}^3\exp\,\bm{\Biggl(}-\frac{(\xi_k-\sum_{l=1}^3R_{kl}u_l)^2}{2c_k}\bm{\Biggr)},
\\
\label{eq:didfusion-after-completing-square}
\tilde{d}_{ij}(\vec{x},\vec{\xi})&=\phantom{-}\frac{4\pi N\kappa^2\Lambda_d}{m\tilde{\Gamma}_\xi}\tilde{\rho}_x(\vec{x})\int\frac{\dd[3]{\vec{u}}}{u^3}\qty[u^2\delta_{ij}-\qty(1-\frac{\zeta}{\Lambda_d})u_iu_j]\prod_{k=1}^3\exp\,\bm{\Biggl(}-\frac{(\xi_k-\sum_{l=1}^3R_{kl}u_l)^2}{2c_k}\bm{\Biggr)},
\end{align}
\end{subequations}
where ${b_i(\vec{x},\vec{p})=\tilde{b}_i\bm{(}\vec{x},\vec{\xi}(\vec{x},\vec{p})\bm{)}}$ and an analogous relation holds
for $\tilde{d}_{ij}$. The substitution $\vec{q}=\mat{R}\vec{u}$ leads to
\begin{subequations}\label{eq:damping-and-diffusion-after-rotation}
\begin{align}
\label{eq:damping-after-rotation}
\tilde{b}_i(\vec{x},\vec{\xi})&=-\frac{8\pi N\kappa^2\Lambda_b}{m\tilde{\Gamma}_\xi}\tilde{\rho}_x(\vec{x})\sum_{j=1}^3R_{ji}\int\dd[3]{\vec{q}}\frac{q_j}{q^3}\prod_{k=1}^3\exp\,\bm{\Biggl(}-\frac{(\xi_k-q_k)^2}{2c_k}\bm{\Biggr)},
\\
\label{eq:didfusion-after-rotation}
\tilde{d}_{ij}(\vec{x},\vec{\xi})&=\phantom{-}\frac{4\pi N\kappa^2\Lambda_d}{m\tilde{\Gamma}_\xi}\tilde{\rho}_x(\vec{x})\sum_{l,m=1}^3R_{li}R_{mj}\int\frac{\dd[3]{\vec{q}}}{q^3}\qty[q^2\delta_{lm}-\qty(1-\frac{\zeta}{\Lambda_d})q_lq_m]\prod_{k=1}^3\exp\,\bm{\Biggl(}-\frac{(\xi_k-q_k)^2}{2c_k}\bm{\Biggr)}.
\end{align}
\end{subequations}
We are not aware of a general representation of the integrals of Eq.~\eqref{eq:damping-and-diffusion-after-rotation} in terms of special functions except for at $\vec{\xi}=\vec{0}$ or in the isotropic case (${c_1=c_2=c_3}$). However, we can reduce the number of integrals from three to one by introducing the representation
\begin{equation}
\label{eq:integral-representation}
\frac{1}{q^3}=\frac{2}{\sqrt{\pi}}\int_0^\infty\dd{s}\sqrt{s}e^{-sq^2},
\end{equation}
which gives
\begin{subequations}\label{eq:damping-and-diffusion-after-representation}
\begin{align}
\label{eq:damping-after-representation}
\tilde{b}_i(\vec{x},\vec{\xi})&=-\frac{16\pi N\kappa^2\Lambda_b}{m\tilde{\Gamma}_\xi}\tilde{\rho}_x(\vec{x})\sum_{j=1}^3R_{ji}\int_0^\infty\dd{s}\sqrt{\frac{s}{\pi}}\int\dd[3]{\vec{q}}q_j\prod_{k=1}^3\exp\,\bm{\Biggl(}-\frac{(\xi_k-q_k)^2}{2c_k}-sq_k^2\bm{\Biggr)},
\\
\label{eq:diffusion-after-representation}
\tilde{d}_{ij}(\vec{x},\vec{\xi})&=\phantom{-}\frac{8\pi N\kappa^2\Lambda_d}{m\tilde{\Gamma}_\xi}\tilde{\rho}_x(\vec{x})\sum_{l,m=1}^3R_{li}R_{mj}\int_0^\infty\dd{s}\sqrt{\frac{s}{\pi}}\int\dd[3]{\vec{q}}\qty[q^2\delta_{lm}-\qty(1-\frac{\zeta}{\Lambda_d})q_lq_m]\prod_{k=1}^3\exp\,\bm{\Biggl(}-\frac{(\xi_k-q_k)^2}{2c_k}-sq_k^2\bm{\Biggr)}.
\end{align}
\end{subequations}
The result of integration over $\vec{q}$ is
\begin{subequations}\label{eq:damping-and-diffusion-after-integration}
\begin{align}
\label{eq:damping-after-integration}
\tilde{b}_i(\vec{x},\vec{\xi})&=-\frac{16\pi^2N\kappa^2\Lambda_b}{m\tilde{\Gamma}_\xi}\tilde{\rho}_x(\vec{x})\sum_{j=1}^3R_{ji}\int_0^\infty\dd{s}\sqrt{s}\frac{\xi_j}{1+2sc_j}\prod_{k=1}^3\sqrt{\frac{2c_k}{1+2sc_k}}\exp(-\frac{s\xi_k^2}{1+2sc_k}),
\\
\label{eq:diffusion-after-integration}
\tilde{d}_{ij}(\vec{x},\vec{\xi})&=\phantom{-}\frac{8\pi^2N\kappa^2\Lambda_d}{m\tilde{\Gamma}_\xi}\tilde{\rho}_x(\vec{x})\sum_{l,m=1}^3R_{li}R_{mj}\int_0^\infty\dd{s}\sqrt{s}\Bigg\{\delta_{lm}\Bigg[\sum_{n \neq l}\frac{1}{1+2sc_n}\qty(c_n+\frac{\xi_n^2}{1+2sc_n})+\frac{\zeta}{\Lambda_d}\frac{1}{1+2sc_l}
\\
\nonumber
&\phantom{=-}\times\qty(c_l+\frac{\xi_l^2}{1+2sc_l})\Bigg]-(1-\delta_{lm})\qty(1-\frac{\zeta}{\Lambda_d})\frac{\xi_l\xi_m}{(1+2sc_l)(1+2sc_m)}\Bigg\}\prod_{k=1}^3\sqrt{\frac{2c_k}{1+2sc_k}}\exp(-\frac{s\xi_k^2}{1+2sc_k}).
\end{align}
\end{subequations}
\end{widetext}

\subsection{Growth rates in the beam frame}
The usual quantity of interest is not the Fokker-Planck coefficients, but rather the growth rates of the distribution's second moments. The relationship between these quantities is given by Eq.~\eqref{eq:growth-rates-FP}. To perform the necessary averaging, we would like to use the distribution of Eq.~\eqref{eq:diagonalized-Gaussian} for $\vec{x}$ and $\vec{\xi}$. For the diffusion term, we simply have ${\langle d_{ij} \rangle = \langle \tilde{d}_{ij}\rangle}$. For the damping term, we must use
\begin{equation}
\langle b_ip_j\rangle = \sum_{l=1}^3R_{lj}\langle \tilde{b}_i\xi_l \rangle - \sum_{l,m=1}^3C^{-1}_{jl}B_{ml}\langle \tilde{b}_i x_m\rangle.
\end{equation} 
As $\tilde{b}_i$ is an even function of $\vec{x}$, the second term vanishes. For the same reason, IBS does not change the mixed momentum-position moments because
\begin{equation}
\dv{t}\langle p_ix_j \rangle = \langle b_ix_j \rangle.
\end{equation}

\begin{figure}
\centering
\includegraphics[scale=0.65]{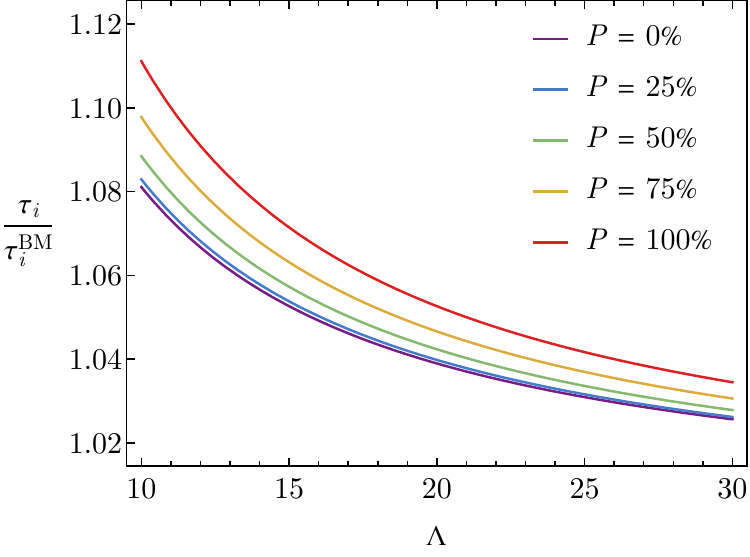}
\caption{\label{fig:GrowthRates}Ratio of the growth times $\tau_i$ including polarization-dependent exchange effects to those calculated by Bjorken and Mtingwa $\tau_i^\mathrm{BM}$ as a function of the Coulomb logarithm $\Lambda$.}
\end{figure}

We will now average Eq.~\eqref{eq:damping-and-diffusion-after-integration} over $\vec{x}$. The required average is a standard Gaussian integral:
\begin{equation}
\label{eq:x-average}
\expval{\tilde{\rho}_x}=\frac{1}{\sqrt{8}\tilde{\Gamma}_x}.
\end{equation}
We thus turn to averaging over $\vec{\xi}$. Before we begin, we note that
\begin{equation}
\expval{\xi_i\xi_j\prod_{k=1}^3\exp(-\frac{s\xi_k^2}{1+2sc_k})} \propto \delta_{ij},
\end{equation}
as for $i \neq j$, the integrand is an odd function of $\xi_i$. The first consequence of this fact is that $\langle\tilde{b}_i\xi_l\rangle$ only contains contributions from terms in Eq.~\eqref{eq:damping-after-integration} with $j=l$. The second consequence is that $\langle\tilde{d}_{ij}\rangle$ only contains contributions from terms in Eq.~\eqref{eq:diffusion-after-integration} with $l=m$. The result of the averaging is then
\begin{widetext}
\begin{subequations}\label{eq:averaged-damping-and-diffusion}
\begin{align}
\label{eq:averaged-damping}
\langle b_i p_j\rangle&=-\frac{16\pi^2 N\kappa^2\Lambda_b}{m\Gamma}\sqrt{c_1c_2c_3}\sum_{k=1}^3R_{ki}R_{kj}c_k\int_0^\infty\dd{s}\frac{\sqrt{s}}{(1+4sc_k)\sqrt{(1+4sc_1)(1+4sc_2)(1+4sc_3)}},
\\
\label{eq:averaged-diffusion}
\langle d_{ij}\rangle&=\phantom{-}\frac{16\pi^2 N\kappa^2\Lambda_d}{m\Gamma}\sqrt{c_1c_2c_3}\sum_{k=1}^3R_{ki}R_{kj}\int_0^\infty\dd{s}\frac{\sqrt{s}}{\sqrt{(1+4sc_1)(1+4sc_2)(1+4sc_3)}}\qty(\sum_{l \neq k}\frac{c_l}{1+4sc_l}+\frac{\zeta}{\Lambda_d}\frac{c_k}{1+4sc_k}).
\end{align}
\end{subequations}
The substitution $s=(4t)^{-1}$ leads to
\begin{subequations}\label{eq:Carlson-damping-and-diffusion}
\begin{align}
\label{eq:Carlson-damping}
\langle b_i p_j\rangle&=-\frac{4\pi^2 N\kappa^2\Lambda_b}{3m\Gamma}\sqrt{c_1c_2c_3}\sum_{k=1}^3R_{ki}R_{kj}c_kR_D(c_l,c_m,c_k),
\\
\label{eq:Carlson-diffusion}
\langle d_{ij}\rangle&=\phantom{-}\frac{4\pi^2N\kappa^2\Lambda_d}{3m\Gamma}\sqrt{c_1c_2c_3}\sum_{k=1}^3R_{ki}R_{kj}\qty[c_lR_D(c_m,c_k,c_l)+c_mR_D(c_k,c_l,c_m)+\frac{\zeta}{\Lambda_d}c_kR_D(c_l,c_m,c_k)],
\end{align}
\end{subequations}
where ${(k,l,m)}$ is a permutation of ${(1,2,3)}$. The appearance of the Carlson elliptic integral $R_D$ in IBS growth rates was noted in Ref.~\cite{Nagaitsev} and the advantage of such a formulation is that $R_D$ can be computed very efficiently~\cite{Carlson}. 

Finally, the growth rates corresponding to Eq.~\eqref{eq:Carlson-damping-and-diffusion} are
\begin{equation}
\label{eq:growth-rates-permutation}
g_{ij}=\frac{4\pi^2N\kappa^2\Lambda_d}{3m\Gamma}\sqrt{c_1c_2c_3}\sum_{k=1}^3R_{ki}R_{kj}\qty[c_lR_D(c_m,c_k,c_l)+c_mR_D(c_k,c_l,c_m)-2c_kR_D(c_l,c_m,c_k)],
\end{equation}
where we have used the identity ${2\Lambda_b-\zeta=2\Lambda_d}$. The Carlson elliptic integrals in Eq.~\eqref{eq:growth-rates-permutation} appear in the special combination $\Psi$ defined in Ref.~\cite{Nagaitsev} as
\begin{equation}
\Psi(x,y,z)=yR_D(z,x,y)+zR_D(x,y,z)-2xR_D(y,z,x).
\end{equation}
Hence, Eq.~\eqref{eq:growth-rates-permutation} can be written as
\begin{equation}
\label{eq:growth-rates}
g_{ij}=\frac{4\pi^2N\kappa^2\Lambda_d}{3m\Gamma}\sqrt{c_1c_2c_3}\qty[R_{1i}R_{1j}\Psi(c_1,c_2,c_3)+R_{2i}R_{2j}\Psi(c_2,c_3,c_1)+R_{3i}R_{3j}\Psi(c_3,c_1,c_2)].
\end{equation}
\end{widetext}
This result is identical to that of Bjorken and Mtingwa except with $\Lambda_d$ in place of $\Lambda$ (see Appendix~\ref{app:BM}). The effect of IBS is often reported using the emittance growth times $\tau_i$, which are defined by
\begin{equation}
\frac{1}{\tau_i}=\frac{1}{\epsilon_i}\dv{\epsilon_i}{t}.
\end{equation}
Each emittance is the average of a quadratic form in the phase-space variables, so the inverse growth times $\tau_i^{-1}$ are linear combinations of the growth rates $g_{ij}$. It follows that
\begin{equation}
\frac{\tau_i}{\tau_i^\mathrm{BM}}=\frac{\Lambda}{\Lambda_d},
\end{equation}
where $\tau_i^\mathrm{BM}$ is the growth time obtained by Bjorken and Mtingwa. This quantity is shown in Fig.~\ref{fig:GrowthRates}. 

\subsection{Growth rates in the lab frame}
The transformation of the growth rates to the lab frame follows Eq.~\eqref{eq:gr-transformation}. However, one should not forget that Eq.~\eqref{eq:bf-Gaussian} is the distribution of phase-space coordinates in the beam frame, and thus both the phase-space volume $\Gamma$ and the eigenvalues $c_i$ are expressed in the beam frame. For example, ${\Gamma = (2\pi p_0)^3\sqrt{\det(\mat{\Sigma})}}$ in terms of the $6 \times 6$ lab-frame  covariance matrix $\mat{\Sigma}$.

\section{Summary}
We have calculated the growth rates due to IBS while taking into account the exchange interaction between identical fermions. Our new growth rates are identical to those calculated by Bjorken and Mtingwa except with a modified Coulomb logarithm. The polarization dependence of the exchange effect is included, and thus our results also indicate the polarization dependence of the growth rates. The influence of the exchange interaction is generally small for typical accelerator parameters, which is expected as particle exchange does not change the small-angle behavior of the cross section. This smallness is in contrast to the Touschek effect, where the polarization dependence is significant because Touschek scattering necessarily involves large angles.

In closing, we note that there are still some relevant calculations to be done in the theory of IBS. For example, it is evident from Eq.~\eqref{eq:damping-and-diffusion-after-integration} that the force resulting from IBS within a Gaussian distribution is a nonlinear function of the coordinates and momenta, and therefore even an initially Gaussian distribution may evolve to a non-Gaussian distribution under the influence of IBS. One can study this effect by numerically solving the Fokker-Planck equation or using Monte Carlo tracking simulations, but some theoretical analysis is also possible, for example using the Hermite expansion suggested for beam-beam simulations by Yokoya in Ref.~\cite{Yokoya-Hermite} (preliminary calculations in this direction were performed in Ref.~\cite{IBS-Hermite}). Another possible calculation is the inclusion of relativistic effects in the beam frame. This subject was treated in Ref.~\cite{Toyomasu}, but only for unpolarized beams. Finally, one can include the polarization-dependent exchange effects in a formalism which takes into account the beam profile and therefore does not contain a logarithmic divergence for small scattering angles. This calculation is performed without exchange effects in Ref.~\cite{Nash-Thesis}, but as this formalism uses the idea of classical trajectories, the modifications required to include quantum effects are not immediately obvious. An approach which includes quantum effects but has not yet been applied to IBS is described in Ref.~\cite{Kotkin-Serbo-Schiller}. Such a calculation  would be very useful as the effects of polarization are likely even clearer in the absence of a (somewhat arbitrary) Coulomb logarithm.

\begin{acknowledgments}
The authors acknowledge many helpful discussions with B.~Nash, D.~T.~Abell, and M.~Blaskiewicz. This work was supported by the U.S. Department of Energy under award number DE-SC0012704. The work of J.~P.~D. was additionally supported by the U.S. Department of Energy, Office of Science, High Energy Physics, under award number DE-SC0024907, the Tigner Traineeships in Accelerator Science program.
\end{acknowledgments}

\appendix 
\section{Connection to Bjorken-Mtingwa results}
\label{app:BM}
The connection to the results of Bjorken and Mtingwa can be shown using Ref.~\cite{Kubo-Oide}, which generalizes the Bjorken-Mtingwa formalism to arbitrary coupling. In Ref.~\cite{Kubo-Oide}, the growth rates are calculated in the frame where $\mat{C}$ is diagonal, in which case the off-diagonal growth rates vanish and the diagonal growth rates are
\begin{equation}
g_{ii} = 2\langle b_ip_i\rangle + \langle d_{ii} \rangle.
\end{equation}
\begin{widetext}
In this case, Eq.~\eqref{eq:damping-and-diffusion-after-completing-square} becomes
\begin{subequations}\label{eq:damping-and-diffusion-diagonalized}
\begin{align}
\label{eq:damping-diagonalized}
\tilde{b}_i(\vec{x},\vec{\xi})&=-\frac{8\pi N\kappa^2\Lambda_b}{m\tilde{\Gamma}_\xi}\tilde{\rho}_x(\vec{x})\int\dd[3]{\vec{u}}\frac{u_i}{u^3}\prod_{k=1}^3\exp\,\bm{\Biggl(}-\frac{(\xi_k-u_k)^2}{2c_k}\bm{\Biggr)},
\\
\label{eq:diffusion-diagonalized}
\tilde{d}_{ii}(\vec{x},\vec{\xi})&=\phantom{-}\frac{4\pi N\kappa^2\Lambda_d}{m\tilde{\Gamma}_\xi}\tilde{\rho}_x(\vec{x})\int\frac{\dd[3]{\vec{u}}}{u^3}\qty[u^2-\qty(1-\frac{\zeta}{\Lambda_d})u_i^2]\prod_{k=1}^3\exp\,\bm{\Biggl(}-\frac{(\xi_k-u_k)^2}{2c_k}\bm{\Biggr)}.
\end{align}
\end{subequations}
The result of averaging over $\vec{x}$ and $\vec{\xi}$ is
\begin{subequations}\label{eq:damping-and-diffusion-diagonalized-averaged}
\begin{align}
\label{eq:damping-diagonalized-averaged}
\langle b_ip_i\rangle&=-\frac{\pi N\kappa^2\Lambda_b}{2m\Gamma}\int\dd[3]{\vec{u}}\frac{u_i^2}{u^3}\prod_{k=1}^3\exp(-\frac{u_k^2}{4c_k}),
\\
\label{eq:diffusion-diagonalized-averaged}
\langle d_{ii}\rangle&=\phantom{-}\frac{\pi N\kappa^2\Lambda_d}{2m\Gamma}\int\frac{\dd[3]{\vec{u}}}{u^3}\qty[u^2-\qty(1-\frac{\zeta}{\Lambda_d})u_i^2]\prod_{k=1}^3\exp(-\frac{u_k^2}{4c_k}),
\end{align}
\end{subequations}
and thus
\begin{equation}
g_{ii} = \frac{\pi N\kappa^2\Lambda_d}{2m\Gamma}\int\dd[3]{\vec{u}}\frac{u^2-3u_i^2}{u^3}\prod_{k=1}^3\exp(-\frac{u_k^2}{4c_k}).
\end{equation}
\end{widetext}
This result differs from Eq.~(59) of Ref.~\cite{Kubo-Oide} only in the replacement of $\Lambda$ by $\Lambda_d$ (${m\alpha^2=\kappa^2/m}$ in units with ${\hbar=c=1}$). 

This computation implies that it most likely would have been simpler to average Eq.~\eqref{eq:damping-and-diffusion-after-rotation} over $\vec{x}$ and $\vec{\xi}$ before reducing the number of integrals using Eq.~\eqref{eq:integral-representation}. However, we reversed the order of these operations to show that the damping vector and the diffusion tensor can be expressed in terms of single integrals [Eq.~\eqref{eq:damping-and-diffusion-after-integration}].

\bibliography{refs.bib}

\end{document}